\documentclass[showpacs,groupedaddress,
preprint,preprintnumbers,amsmath,amssymb,floatfix]{revtex4}
\usepackage{amsmath,graphicx,amsfonts}
\usepackage{dcolumn}

\newcommand{\be}{\begin{equation}}
\newcommand{\ee}{\end{equation}}

\begin{document}

\title{Conformational transformations induced by the
charge-curvature interaction at finite temperature}

\author{Yu. B. Gaididei}
\affiliation{Bogolyubov Institute for Theoretical Physics,
Metrologichna str. 14 B, 01413, Kiev, Ukraine}
\author{C. Gorria}
\affiliation{Department of Applied Mathematics and Statistics,
University of the Basque Country, E-48080 Bilbao, Spain  }
\author{P. L. Christiansen}
\affiliation{Informatics and Mathematical Modeling and Department of
Physics, The Technical University of Denmark, DK-2800 Lyngby,
Denmark}
\author{M.P. S{\o}rensen}
\affiliation{ Department of
Mathematics, The Technical University of Denmark, DK-2800 Lyngby,
Denmark}

\date{ }

\begin{abstract}
The role of thermal fluctuations on the conformational dynamics of
a single closed filament is studied. It is shown that, due to the
interaction between charges and bending degrees of freedom,
initially circular aggregates may undergo transformation to
polygonal shape. The transition occurs both in the case of
hardening and softening charge-bending interaction. In the former
case the charge and curvature are smoothly distributed along the
chain while in the latter spontaneous kink formation is initiated.
The transition to a non-circular conformation is analogous to the
phase transition of the second kind.
\end{abstract}


\maketitle

\section{Introduction}

Conformational flexibility is a fundamental property of biological
systems which determines their functioning \cite{shape,geom,doi}.
Even modest conformational changes modify long-range electronic
interactions in oligopeptides \cite{wolfgang}, they may  remove
steric hindrances and open the pathways for  molecular motions which
are not available in rigid proteins \cite{feitel}. The DNA
conformation in the nucleosome core is crucial for gene replication,
transcription and recombination \cite{richmon}. Recent DNA
cyclization experiments \cite{cloutier,vologodskii,yuan} have shown
the facile {\it in vitro} formation of DNA circles shorter than 30
nm (100 base pairs) which is even shorter than commonly accepted
persistence length $50 nm$ ($150$ base pairs). This means that the
worm-like chain model does not work for such short DNA
molecules and to explain this phenomenon one should allow local
softenings of DNA which facilitates disruptions (kinks) in the
regular DNA structure \cite{du,yan,wiggins}. According to \cite{du}
the kink formation is due to strong DNA bending while in \cite{yuan}
that the softening originates from Watson-Crick base-pair breathing.
An alternative approach which allows to avoid  kinking was
proposed in Ref. \cite{nelson} where a class of models with
nonlinear DNA elasticity was introduced. It was shown in
\cite{nelson} that a "subelastic chain" model, in the frame of which
the bending energy is proportional to absolute value of curvature,
can reproduce the main features of Cloutier and
Widom's experiments \cite{cloutier}.

Quite recently a simple, {\it generic} model for charge-curvature
interactions on closed  molecular aggregates was proposed
\cite{gaid-conf}. It was shown that the presence of charge
modifies (softens or hardens) the local chain stiffness. It was
found that due to the interaction between charge carriers and the
bending degrees of freedom the circular shape of the aggregate may
become unstable and the aggregate takes the shape of an ellipse
or, in general, of a polygon. It was shown also that when the
charge-curvature interaction leads to softening the local chain
stiffness kinks spontaneously appear in the chain.

These results were obtained by using  the mean-field approach where
thermal fluctuations are ignored  and strictly speaking, this
approach is valid only for zero-temperature. In the case of finite
temperature the interaction with environment and thermal
fluctuations have to be considered.

The aim of this paper is to extend the  results of Ref.
\cite{gaid-conf} to the case of finite temperature. We study the
charge-induced conformational  transformations of closed molecular
aggregates in the presence of thermal fluctuations which we model
in the frame of Langevin dynamics. The paper is organized as
follows. In Sec. II we describe a model. In Sec. III we present an
analytical approach to the problem. In Sec. IV we display
the results of numerical simulations and compare with the analytical
results. In Sec. V we discuss some concluding remarks.


\section{The model}

We consider a  polymer chain consisting of L units ( for DNA each
unit is a base pair) labelled by an index $l$, and located at the
points $\vec{r}_l=(x_l, \, y_l), ~l=1\dots L$. We are
interested in the case when the chain is closed, therefore we impose
the periodicity condition on the coordinates,
\begin{equation}\label{closed}
\vec{r}_l=\vec{r}_{l+L}.
\end{equation}
We assume that there is a small amount of mobile carriers
(electrons, holes in the case of DNA, protons in the case of
hydrogen bonded systems) on the chain. The Hamiltonian of the
system can be presented as the sum
\begin{equation}\label{hamilt-tot}
H_{tot}=H+H_{stoch}.
\end{equation}
The first term in this equation is the Hamiltonian of an isolated
filament introduced in Ref. \cite{gaid-conf}
\begin{equation}\label{hamilt}
H=U_b+U_s+H_{el}+H_{el-conf}.
\end{equation}
Here
\begin{equation}\label{potbend}
U_b=\frac{k}{2}\,\sum_l\,\frac{\kappa^2_l}{1-\kappa_l^2/\kappa_{max}^2}
\end{equation}
is the bending energy of the chain where
\begin{equation}\label{kappan}
\kappa_l\equiv|\vec{\tau}_{l}-\vec{\tau}_{l-1}|=2\,\sin\frac{\alpha_l}{2}
\end{equation}
determines the curvature of the chain at the point $l$. Here
\begin{equation}\label{tau}\vec{\tau}_{l} =
\frac{\vec{r}_{l+1}-\vec{r}_l}{|\vec{r}_{l+1}-\vec{r}_l|}
\end{equation}
is the tangent vector at the point $l$ of the chain and $\alpha_l$
is the angle between the tangent vectors $\vec{\tau}_{l}~$ and
$~\vec{\tau}_{l-1}$, $k$ is the elastic modulus of the bending
rigidity (spring constant)  of the chain. The term
$\kappa_l^2/\kappa_{max}^2$ in Eq. (\ref{potbend}) gives the
penalty for too large bending deformations. Here the parameter
$\kappa_{max}=2\,\sin\left(\alpha_{max}/2\right)$ is the maximum
local curvature with $\alpha_{max}$ being the maximum bending
angle. The second term in Eq. (\ref{hamilt})
\begin{equation}\label{potstr}
U_s=\frac{\sigma}{2}\,\sum_l\left(|\vec{r}_l-\vec{r}_{l+1}|-a\right)^2
\end{equation}
determines the stretching energy with $\sigma$ being an elastic
modulus of the stretching rigidity of the chain and $a$ is the
equilibrium distance between units (in what follows we assume
$a=1$). We take the simplest theoretical model for charge
carriers, a nearest neighbor tight binding Hamiltonian in the form
\begin{equation}
\label{hopp} H_{el}=J\,\sum_{l}\Big |
\psi_l - \psi_{l+1}\Big |^2 \; ,
\end{equation}
where $\psi_l$ is the  wave function of carrier localized at
$r_l$ and $J$ measures the carrier hopping between adjacent sites.
The last term in Eq. (\ref{hamilt}) represents the
charge-curvature interaction. In the small curvature limit it has
the form
\begin{equation}\label{el-conf}
H_{el-conf}=-\frac{1}{2}\,\sum_{l}\chi\,|\psi_l|^2\,\left(\kappa_{l+1}^2+
\kappa_{l-1}^2\right),
\end{equation}
here $\chi$ is the coupling constant. Combining Eqs.
(\ref{potbend}) and (\ref{el-conf}), we notice that the effective
bending rigidity changes close to the points where the electron
(hole) is localized. For positive values of the coupling constant
$\chi$ there is a local softening of the chain, while for $\chi$
negative there is a local hardening of the chain.

The quantity
\begin{equation}\label{density}
\nu\equiv\frac{1}{L}\sum_l\,|\psi_l|^2
\end{equation}
gives the total density of charge carriers which can move along the
chain and participate in the formation of the conformational state
of the system. The second term in Eq. (\ref{hamilt-tot}) describes
the interaction of the filament with fluctuating environment
\begin{equation}\label{intstoch}
  H_{stoch}=\sum_l\,\vec{r}_l\cdot \vec{R}_l(t)
\end{equation}
where the stochastic forces
$\vec{R}_l(t)=\left(X_l(t),~Y_l(t)\right)$ are the Gaussian white
noise
\begin{equation}\label{noise}
\begin{array}{l}
\langle X_l(t)\rangle=\langle Y_l(t)\rangle=0, \\
\langle X_l(t)\,X_{l'}(t')\rangle=
 \langle Y_l(t)\,Y_{l'}(t')\rangle=
 2\,D\,\delta_{l\,l'}\,\delta(t-t'), \\
 \langle X_l(t)\,Y_{l'}(t')\rangle=0
\end{array}
\end{equation}
with $D$ the standard deviation.

To analyze the evolution of the shape of the filament, it is
convenient to introduce the radius-of-gyration tensor ${\bf I}$ as
in Ref. \cite{solc,rudnick}. Its components are
\begin{eqnarray}\label{moment}
{\bf
I}_{xx}(t)=\frac{1}{L}\,\sum_{l}\left(x_{l}(t)-x^c(t)\right)^2,\nonumber\\
{\bf
I}_{yy}(t)=\frac{1}{L}\,\sum_{l}\left(y_{l}(t)-y^c(t)\right)^2,\nonumber\\
{\bf
I}_{xy}(t)=\frac{1}{L}\,\sum_{l}\left(x_{l}(t)-x^c(t)\right)\left(y_{l}(t)-y^c(t)\right)
\end{eqnarray}
where
\begin{equation}\label{cm}
\left(x^c(t),y^c(t)\right)=\frac{1}{L}\,\sum_l\left(x_l(t),y_l(t)\right)
\end{equation}
is the center-of-mass coordinate. The square roots of the two
eigenvalues $R_{q}=\sqrt{{\bf I}_q},~q=1,2$ of the tensor ${\bf
I}$ give the two principal radii of the system. They express the
sizes of the filament along the major and minor axis. As it is seen
from Eqs. (\ref{moment}) the eigenvalues have the form
\begin{equation}\label{dev}
{\bf I}_{1,2}=\frac{1}{2}\left[\left({\bf I}_{xx}+{\bf
I}_{yy}\right)\pm\sqrt{\left({\bf I}_{xx}-{\bf
I}_{yy}\right)^2+4\,{\bf I}_{xy}^2}\,\right].
\end{equation}
Indexes 1 and 2 correspond to the $+$ and $-$ sign respectively. To
characterize the shape of conformation it is convenient to introduce
the quantity
\begin{equation}\label{aspherity}
A = {\bf I}_1-{\bf I}_2\equiv \sqrt{\left({\bf I}_{xx}-{\bf
I}_{yy}\right)^2+4\,{\bf I}_{xy}^2},
\end{equation}
defined as the ``aspherity'' \cite{rudnick}. It characterizes the
shape's overall deviation from circular symmetry which corresponds
to $A=0$.

\section{Analytic approach}

The aim of this section is to develop an analytical approach, which
provides a better insight into the physical mechanism of
conformational transformations induced by the charge-curvature
interaction in the fluctuating media. We will assume that  the
characteristic size of the excitation is much larger than the
lattice spacing and  replace $\psi_l$ and $\vec{r}_l$ by the
functions $\psi(s,t)$ and $\vec{r}(s,t)$, respectively. Here the
arclength $s$ is the continuum analogue of $l$. We assume that
the chain is inextensible and this assumption is expressed by the
constraint
\begin{equation}\label{inextens}
|\partial_s\vec{r}|^2=1.
\end{equation}
which is automatically taken into account by choosing the parametrization
\begin{equation}\label{param}\partial_s\,x(s)=\sin\theta(s),
~~~\partial_s\,y(s)=\cos\theta(s)
\end{equation}
where the angle $\theta(s)$ satisfies the periodicity condition of
Eq. (\ref{closed})
\begin{equation}\label{period-pos}
\theta(s+ L)=2\pi +\theta(s),
\end{equation}
\begin{equation}\label{closure}
\int\limits_0^L\,\cos\,\theta(s)\,ds=\int\limits_0^L\,\sin\,\theta(s)\,ds=0.
\end{equation}
In the frame of the parametrization (\ref{param}) the shape of the chain
is determined by the equations
\begin{equation}\label{shape}
x(s)=\int\limits_0^s\,\sin\,\theta(s')\,ds',~~~
y(s)=\int\limits_0^s\,\cos\,\theta(s')\,ds'.
\end{equation}
In the continuum limit the curvature (\ref{kappan}) takes
the form $\kappa(s)= |\partial^2_s \vec{r}(s)|$, which is given by
\begin{equation}\label{kappas}
\kappa(s)=\partial_s \theta.
\end{equation}
The continuum version of the total Hamiltonian of the system can be
written as the sum
\begin{equation}\label{totalham}
H_{tot}=H+H_{stoch}(t)
\end{equation}
where
\begin{equation}\label{energy}
H=\int\limits_{0}^{L} \Biggl\{ J\, |\partial_s\psi |^2
\,+\,\left(\frac{k}{2}-
\chi\,|\psi^2|\right)\,\left(\partial_s\theta\right)^2 \Biggr\} ds
\end{equation}
is the analogue to the Hamiltonian (\ref{hamilt}) and
\begin{equation}\label{stochast}
H_{stoch}(t)=\int\limits_{0}^{L}\left(X(s,t)\,x(s,t)+Y(s,t)\,y(s,t)\right)\,ds
\end{equation}
gives the interaction of the chain with the fluctuating environment
(\ref{intstoch}) in the continuum limit.  The stochastic forces
$\vec{R}(s,t)=\left(X(s,t),\,Y(s,t)\right)$ are the continuum
version of the forces $X_l(t)$ and $Y_l(t)$. They obey the relations
\begin{eqnarray}\label{stoch-cont}
\langle X(s,t)\,X(s',t')\rangle= \langle Y(s,t)\,Y(s',t')\rangle =
\nonumber\\
2 D\,\delta(s-s')\,\delta(t-t'),\nonumber\\
\langle X(s,t)\,Y(s',t')\rangle=0.
\end{eqnarray}
We will restrict  our analysis to the case when the filament shape
only slightly deviates from the circle. Therefore in the derivation
of the Hamiltonian (\ref{energy}) we could neglect the term
$\kappa^2/\kappa_{max}^2$ in the denominator of Eq. (\ref{potbend}).

By using the Madelung transformation
\begin{equation}\label{madelung}\psi(s,t)=\sqrt{\rho(s,t)}\,e^{i\,\phi(s,t)}
\end{equation}
where $\rho(s,t)$ is the charge density and $\phi(s,t)$ is  the
phase, the Hamiltonian (\ref{energy}) can be written as follows
\begin{equation}\label{energy-rho}
H=\int\limits_{0}^{L} \Biggl\{ J\, \Biggl[\frac{\left(\partial_s\rho
\right)^2}{4\,\rho}\, \,+\rho\,\left(\partial_s
\phi\right)^2\Biggr]+\,\left(\frac{k}{2}-
\chi\,\rho\right)\,\left(\partial_s\theta\right)^2 \Biggr\} d s.
\end{equation}

The dynamics of the system is governed by the Hamilton equations for
the charge variables $\rho(s,t)$, $\phi(s,t)$
\begin{equation}\label{var-pr-1}
\frac{\delta{\cal L}}{\delta \rho}=0 \qquad \textrm{and} \qquad
\frac{\delta{\cal L}}{\delta \phi}=0
\end{equation}
where
\begin{equation}\label{lagrangian}
{\cal L}=-\,\int\limits_0^L\,\rho\,\partial_t\phi\,ds-H
\end{equation}
is the Lagrangian of the system and $\delta/\delta(\cdot)$ is a
variational derivative. By introducing the dissipation function
\begin{equation}\label{dissip}
{\cal
F}=\eta\,\frac{1}{2}\,\int\limits_0^L\,\left(\partial_t\vec{r}\right)^2\,ds
\end{equation}
the Langevin equation for the position $\vec{r}(s,t)$ can also be
written in the variational form
\begin{equation}\label{var-pr-2}
\frac{\delta {\cal F}}{\delta\partial_t\vec{r}(s,t)}= -\frac{\delta
\,H_{tot}}{\delta\,\vec{r}(s,t)}.
\end{equation}
Eqs. (\ref{var-pr-1}) and (\ref{var-pr-2}) should be considered with
the periodicity condition (\ref{period-pos}) and the closure
condition (\ref{closure}) for the position $\vec{r}(s,t)$. Now the
periodicity conditions for the charge variables take the form
\begin{equation}\label{period-charge}
\rho(s)=\rho(s+L),\qquad \phi(s)=\phi(s+L)
\end{equation}
and the normalization condition in (\ref{density}) becomes
\begin{equation}\label{norm-cond}
\frac{1}{L}\,\int\limits_0^L\,\rho\,ds=\nu.
\end{equation}
To take into account the periodicity conditions (\ref{period-pos}) and
(\ref{period-charge}) we can expand the curvature and the charge
variables in the Fourier series:
\begin{eqnarray}
& \displaystyle \partial_s\theta(s,t)=\frac{2 \pi}{L} \,
\left[1+\,\sum\limits_{j\ge 2}c_j(t)\,\cos\left(\frac{2\,\pi
j\,s}{L}\right)\right], \label{fourier-curv}\\
& \displaystyle \rho(s,t)= \nu\, \,\left[1+\,\sum\limits_{j\ge 2}\zeta_j(t)\,
\cos\left(\frac{2\,\pi j\,s}{L}\right)\right], \label{fourier-dens}\\
& \displaystyle \phi(s,t)=\,\sum\limits_{j\ge 2}\Phi_j(t)\,\cos\left(\frac{2\,\pi
j\,s}{L}\right). \label{fourier-phase}
\end{eqnarray}
Note that the first harmonic with $j=1$ does not contribute to the
Fourier expansion (\ref{fourier-curv}) due to the closure condition
(\ref{closure}). The coefficient $\nu$ in the expansion
(\ref{fourier-dens}) takes into account the normalization condition
(\ref{norm-cond}). Different harmonics in the Fourier expansion
(\ref{fourier-curv}) represent different types of shape deformation.
For example the term with $j=2$ determines an elliptic deformation,
the term with $j=3$ represents a triagonal deformation, etc (see
\cite{gaid-conf} for more detail). For the sake of simplicity we
consider only the elliptic deformation of the filament ($j=2$) and
restrict ourselves to the case when the deviations from the circular
shape are small and the charge distribution along the chain is
smooth: $|c_2|,|\rho_2|,|\phi_2|\,<\,1,~|c_j|,|\rho_j|,|\phi_j|\,\ll\,1~$ for
$j\geq 3$. Thus the expansions
(\ref{fourier-curv})-(\ref{fourier-phase}) reduce to
\begin{eqnarray}
& \displaystyle \partial_s\theta(s,t)=\frac{2\pi}{L}\,
\left[1+c(t)\,\cos\left(\frac{4\pi\,s}{L}\right)\right], \label{fourier-reduce1}\\
& \displaystyle \rho(s,t)=\nu\,\left[1+\zeta(t)\,\cos\left(\frac{4\pi\,s}{L}\right)\right], \label{fourier-reduce2}\\
& \displaystyle \phi(s,t)=\Phi(t)\,\cos\left(\frac{4\pi\,s}{L}\right) \label{fourier-reduce3}
\end{eqnarray}
where we omitted subscript in the notations for the Fourier
harmonics.

Inserting Eqs.
(\ref{fourier-reduce1})-(\ref{fourier-reduce3}) into
Eqs. (\ref{stochast}) , (\ref{energy-rho}), (\ref{lagrangian}) and
(\ref{dissip}) we get
\begin{equation}\label{lagrangian-eff}
{\cal L}_{eff}=-\frac{L}{2}\,\nu\,\zeta\,\frac{d\Phi}{dt}-{\cal H}.
\end{equation}
Here
\begin{equation}\label{hamiltonian-eff}
{\cal H} = \frac{\pi^2}{L}\left(-4\,J\,\nu\,\sqrt{1-\zeta^2}+
8\,J\,\nu\,\Phi^2 + \,k_e\,c^2-4\,\chi\,\nu\,\zeta\,c\right)
\end{equation}
is the effective Hamiltonian with some irrelevant constants being
omitted. In Eq. (\ref{hamiltonian-eff}) $k_e=k-2\,\chi\,\nu$ is an
effective bending rigidity of the filament for the case when the
charge is unformly distributed along the chain.
\begin{equation}\label{stoch-eff}
\begin{array}{c}
\displaystyle {\cal
H}_{stoch}(t)=\frac{L^2}{4\pi^2}\int\limits_{0}^{2\pi} \left[\,
X\left(\frac{L\,s}{2\pi},t\right) \right. \\ \displaystyle
\int\limits_0^{s}\cos\left(s'-\frac{c}{2}\,
\sin\left(2\,s'\right)\right)\, ds' \\ \displaystyle \left.
+Y\left(\frac{L\,s}{2\pi},t\right) \int\limits_0^{s}
\cos\left(s'+\frac{c}{2}\sin\left(2\,s'\right)\right)\, ds'
\right]\, ds
\end{array}
\end{equation}
is the effective interaction with stochastic forces, and
\begin{equation}\label{dissip-eff}
{\cal F}=\frac{1}{2}\,b(c)\,\left(\frac{d\,c}{dt}\right)^2
\end{equation}
is the effective dissipative function. The damping coefficient
$b(c)$ has the form
\begin{equation}\label{bkappa}
b(c)=\,\frac{1}{4}\,\eta\,\frac{L^3}{(2\pi)^3}\,\int\limits_{0}^{2\pi}\,
\left(\alpha^2(c,s)+\alpha^2(-c,s)\right)\,ds
\end{equation}
where the notation
\begin{equation}\label{akappa}
\alpha(c,s)=\int\limits_0^s\,\sin(2\,s')\,
\sin\left(s'+\frac{c}{2}\,\sin(2\,s')\right)\,ds'
\end{equation}
is introduced. Note also that in the derivation of Eq.
(\ref{stoch-eff}) we took into account the periodicity in the
stochastic terms, $X(s+L,t)=X(s,t)$ and $Y(s+L,t)=Y(s,t)$.

Equations of motion for the quantitites $\Phi$, $\zeta$ and $c$
follow from Eqs. (\ref{energy-rho}), (\ref{var-pr-1}),
(\ref{lagrangian-eff})-(\ref{dissip-eff}) and they have the form
\begin{equation}\label{eqphi}
\frac{d\,\Phi}{dt}=-\frac{8\pi^2}{L^2}
\left(J\,\frac{\zeta}{\sqrt{1-\,\zeta^2}}-\chi\,c\right),
\end{equation}
\begin{equation}\label{eqzeta}
\frac{d\,\zeta}{dt}=\frac{32\,\pi^2}{L^2}\,J\,\Phi,
\end{equation}
\begin{equation}\label{eqkappa}
\frac{d\,c}{dt}=-\frac{2\,\pi^2}{b(c)\,L}\,
\Biggl(k_e\,c-2\,\chi\,\nu\,\zeta\Biggr)- f(c,t)
\end{equation}
where
\begin{eqnarray}\label{stoch-force}
f(c,t)= \frac{L^2}{4\,b(c)\,\pi^2}\,
\int\limits_0^{2\pi}\left[X\left(\frac{L\,s}{2\,\pi},t\right)\,\alpha(-c,s)
\right. \nonumber\\ - \left.
Y\left(\frac{L\,s}{2\,\pi},t\right)\,\alpha(c,s)\right]\, ds.
\end{eqnarray}
is an effective stochastic force.

Note that, in terms of the Ansatz (\ref{fourier-reduce1}), the
aspherity $A$, which is defined by Eq. (\ref{aspherity}), can be
written approximately as
\begin{equation}\label{aspherity-anal}
A=\,\frac{L^2}{8\,\pi^2}\,c.
\end{equation}

Let us analyze the cases of zero temperature and finite temperature
separately.


\subsection{Deterministic behavior: zero temperature limit}

In the no-noise case the dynamics of the system is described by Eqs.
(\ref{eqphi})-(\ref{eqkappa}) with $X=Y=0$.
The system under consideration is characterized by the
control parameter
\begin{eqnarray}\label{xi}\xi=\frac{\Delta_{def}}{\Delta_{disp}}
\end{eqnarray}
which is the ratio of the deformation energy (i.e. the  energy shift
due to the charge-bending interaction)
\begin{eqnarray}\label{def}\Delta_{def}=
\left(\frac{2\,\pi}{L}\right)^2\,\nu^2\,
\frac{\chi^2}{k_e}\end{eqnarray}
with respect to the dispersion energy
\begin{eqnarray}\label{disp}\Delta_{disp}=\frac{1}{2}\,
\left(\frac{2\,\pi}{L}\right)^2\,\nu\,J.\end{eqnarray}
A simple analysis shows that when the charge-curvature coupling is weak
such that the control parameter, $\xi<1$, these equations have a unique
stationary point $\Phi=0$, $\zeta=0$ and $c=0$. This state corresponds to
a uniformly distributed charge along the circular filament.

When $\xi\,>\,1$ there are two equivalent stationary states
\begin{eqnarray}\label{stationary-det}
\Phi=0,~~\zeta_{\pm}=\pm\,\sqrt{1- \frac{1}{\xi^2}},
~~c_{\pm}=\frac{2\,\chi\,\nu}{k_e}\,\zeta_{\pm}
\end{eqnarray}
which represent an elliptically deformed filament with a spatially
non-uniformly distributed charge. The two solutions $c_{\pm}$
correspond to two mutually orthogonal directions in which the
filament may be elongated. Note that in the case of softening
charge-curvature interaction ($\chi>0$) the maxima of the curvature
and the charge density coincide while in filaments with hardening
charge-curvature interaction ($\chi<0$) the curvature of the
filament is minimal (the filament is locally more flat) in the
places where the charge density is maximal.

\subsection{Charge-charge correlation effects}

The aim of this subsection is to clarify the role of interaction
between charge carriers in the formation of polygonally shaped
aggregates in the zero-temperature limit. In describing the
charge-charge repulsion effects, we will use a on-site  interaction
in the form
\begin{eqnarray}\label{A1}
H_{el-el}=\frac{1}{2}\,V\,\sum_{n}\,|\psi_n|^4\end{eqnarray}
which in the continuum limit in terms of charge variables (\ref{madelung})
reads
\begin{eqnarray}\label{A2}
H_{el-el}=
\frac{1}{2}\,V\,\int\limits_{0}^{L}\,\rho^2\,ds.\end{eqnarray}
The parameter $V$ in Eqs. (\ref{A1}) and (\ref{A2}) characterizes the
strength of the interaction. Thus the Hamiltonian of the system with
account of charge-charge interaction effects has the form
\begin{eqnarray}\label{A3}
H_{cc}=H+\frac{1}{2}\,V\,\int\limits_{0}^{L} \rho^2
 d s.
\end{eqnarray}
where the Hamiltonian $H$ is given by Eq. (\ref{energy-rho}).
Inserting in Eq. (\ref{A3}) the Ansatz (\ref{fourier-reduce1}), we
get
\begin{eqnarray}\label{A4}
{\cal H}_{cc} = \frac{\pi^2}{L}\Biggl(-4\,J\,\nu\,\sqrt{1-\zeta^2}+
\frac{L^2}{4\,\pi^2}\,V\,\nu^2\,\zeta^2+ \nonumber\\
8\,J\,\nu\,\Phi^2 + \,k_e\,c^2-4\,\chi\,\nu\,\zeta\,c\Biggr).
\end{eqnarray}
An inspection of the function (\ref{A4}) shows that in the
zero-temperature limit the spatially uniform charge distribution
along the circular filament becomes unstable for the control
parameter (\ref{xi}) satisfying the inequality
\begin{eqnarray}\label{A5}\xi\,>1+\frac{L^2}{8\,\pi^2}\,
\nu\,\frac{V}{J}.\end{eqnarray}
Thus  the elliptic shape more easily
arises in short filaments with  strong charge-bending interaction
$\chi$ and  relatively weak interaction between charge carriers
$V$.

In what follows   we will assume that  the charge-bending
interaction is strong, $\Delta_{def} \gg \nu^2\,V/4$,
and for the sake of simplicity  will neglect the interaction between
charges.


\subsection{Stochastic behavior: finite temperature}

We will study the role of thermal fluctuations by using a formalism
of the Fokker-Planck equation. To this end we introduce the
probability distribution density
\begin{equation}\label{prob-density}
P\left(c,\zeta,\Phi;t\right) =
\Big\langle\delta\left(c-c(t)\right)\,
\delta\left(\zeta-\zeta(t)\right)\,\delta\left(\Phi-\Phi(t)\right)\Big\rangle
\end{equation}
As it is seen from Eqs. (\ref{stoch-cont}) and (\ref{stoch-force})
the stochastic forces $f(c,t)$ represent Gaussian white noise with
the mean value
\begin{equation}\label{mean}
\langle f(c,t)\rangle=0
\end{equation}
and the two time covariance given by
\begin{equation}\label{covar}
\langle f(c,t)\,f(c',t')\rangle=2\,F(c,c')\,\delta(t-t')
\end{equation}
where
\begin{eqnarray}\label{cov-F}
F(c,c')=\,\frac{D}{b(c)\,b(c')}\,
\frac{L^3}{8\,\pi^3}\,\int\limits_0^{2\pi}\Big(\alpha(-c,s)\,\alpha(-c',s)
\nonumber\\ + \alpha(c,s)\,\alpha(c',s)\Big)\,ds.
\end{eqnarray}
It is straightforward to obtain (see e.g. \cite{gardiner})  that the
Fokker-Planck equation which describes the time evolution of the
probability distribution (\ref{prob-density}) of a set of Langevin
equations (\ref{eqphi})-(\ref{eqkappa}) in the
Stratonovich sense  has the form
\begin{eqnarray}\label{FP}
\displaystyle
\partial_t\,P=-\partial_{\zeta}\left(P\,\partial_{\Phi}{\cal
H}\right) +\partial_{\Phi}\left(P\,\partial_{\zeta}{\cal H}\right)+
\partial_{c}\left(\frac{P}{b(c)}\,\partial_{c}{\cal H}\right) \nonumber\\
\displaystyle -\partial_{c}\left[P\,\left(\partial_{c}F(c,c')\right)
\Big|_{c'=c}\right]+\
\partial_{c}^2\left(P\,F(c,c)\right) \qquad
\end{eqnarray}
where the Hamiltonian ${\cal H}$  is given by Eq.
(\ref{hamiltonian-eff}) Inserting into Eq. (\ref{FP}) the relations
\begin{equation}\label{Fb}
F(c,c)=\frac{T}{b(c)}, ~~
\partial_{c}F(c,c')\Big |_{c'=c}=
\frac{T}{2}\frac{d\,}{d\,c}\left(\frac{1}{b(c)}\right)
\end{equation}
which follow from Eqs. (\ref{bkappa}), (\ref{akappa}) and
(\ref{cov-F}), we obtain  the equation for the probablity
distribution $P\left(c,\zeta,\Phi;t\right)$ in the form
\begin{eqnarray}\label{FPeq}
\displaystyle
\partial_t\,P=-\partial_{\zeta}\left(P\,\partial_{\Phi}{\cal
H}\right) +\partial_{\Phi}\left(P\,\partial_{\zeta}{\cal H}\right)+
\partial_{c}\left(\frac{P}{b(c)}\,\partial_{c}{\cal H}\right) \nonumber\\
\displaystyle -\frac{T}{2}\,\partial_{c}\left[
P\,\frac{d\,}{dc}\left(\frac{1}{b(c)}\right)\right]+
T\,\partial_{c}^2\left(\frac{1}{b(c)}\,P\right). \qquad
\end{eqnarray}

It is interesting to note that if the
stochastic force $f(c,t)$ in the Langevin
equations (\ref{eqphi})-(\ref{eqkappa}) is replaced by
\begin{equation}\label{stoch-force-mod}
f_{mod}=\frac{1}{\sqrt{b(c)}}\,\xi(t)
\end{equation}
where $\xi(t)$ is a white noise with
\begin{equation}\label{wn}
\langle\xi(t)\rangle=0,~~~\langle\xi(t)\,\xi(t')\rangle=2\,D\,\delta(t-t').
\end{equation}
We obtain the same Fokker-Planck relation for this new set of
Langevin equations. This means that the stochastic equations
(\ref{eqphi})-(\ref{eqkappa}) choosing $f(c,t)$
from Eqs (\ref{stoch-cont}) and (\ref{stoch-force}) or from Eqs.
(\ref{stoch-force-mod}) and (\ref{wn}) are equivalent.

The  stationary probability distribution (i.e. the solution of the
Fokker-Planck equation (\ref{FPeq}) for $t\rightarrow\infty$) is
given by
\begin{equation}\label{stationary}
P_{st}=C\,\sqrt{b(c)}\,e^{-{\cal H}/T}
\end{equation}
where $C$ is a normalization constannt.

Assuming that $c\,<\,1$, one can expand the function $a(c,s)$ from
Eq. (\ref{akappa}) into a Taylor series and keeping only leading
terms one can obtain that
\begin{equation}\label{damping-coeff}
\ln b(c)\approx const+0.04 \,c^2.
\end{equation}
Combining Eqs. (\ref{hamiltonian-eff}), (\ref{stationary}) and
(\ref{damping-coeff}) we obtain that the effective bending rigidity
({\it i.e.} the coefficient in front of $c^2$ in the expression
${\cal H}+\frac{1}{2}\,\ln b(c)$) becomes temperature dependent and
takes the form $k_e-0.002 L\,T$. We will assume that the persistence
length $l_p=k_e/T$  satisfies the inequality $l_{p}\gg 0.002 L$  and
in what follows we will neglect the temperature dependence of the
effective bending rigidity.

By integrating the stationary probability distribution
(\ref{stationary}) over the charge variables $\Phi$ and $\zeta$
(note that $\zeta\in \left(-1,1\right)$  because under definition
the charge density $\rho(s,t)>0$) we obtain a reduced distribution
\begin{equation}\label{reduced}
\begin{array}{c}
\displaystyle {\cal P}(c)=\frac{1}{{\cal N}}\,
\int\limits_{0}^{\pi/2}\,d\theta
\,\sin\theta\,e^{\beta\,\sin\theta}\,
\cosh\left(\beta\frac{\chi}{J}\,c\,\cos\theta\right) \\
\displaystyle
\exp\left[-\frac{\beta}{2\,\xi}\,\left(\frac{\chi}{J}\right)^2\,
c^2\right],
\end{array}
\end{equation}
where
\begin{equation}\label{norm-const}
\begin{array}{c}
\displaystyle {\cal N}=
\int\limits_{-\infty}^{\infty}d\,c\,\int\limits_{0}^{\pi/2}\,d\theta
\,\sin\theta\,e^{\beta\,\sin\theta}\,\cosh\left(\beta\frac{\chi}{J}\,c\,\cos\theta\right)
\\
\displaystyle
\exp\left[-\frac{\beta}{2\,\xi}\,\left(\frac{\chi}{J}\right)^2\,c^2\right]
\end{array}
\end{equation}
is the normalization constant. In Eqs. (\ref{reduced}) and
(\ref{norm-const}) $\beta \,=4\,\pi^2\,J\,\nu/(L\,T)$ is a
dimensionless inverse temperature. The function ${\cal P}(c)$ gives
the probability of finding the curvature $c$ in the interval
$(c,c+d\,c)$ irrespective of the magnitude of the charge variables.

\begin{figure}
\includegraphics[width=8.cm]{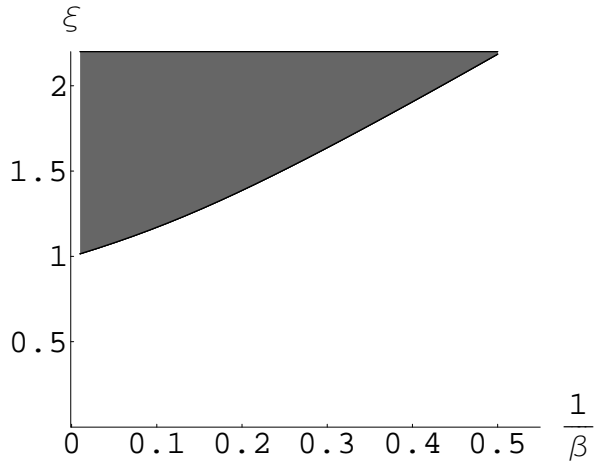}
\caption{\label{figphase} The probability distribution phase
diagram. In the shaded area the probability distribution is
bi-modular and in the unshaded area it is unimodular.}
\end{figure}

\begin{figure}
\includegraphics[width=8.cm]{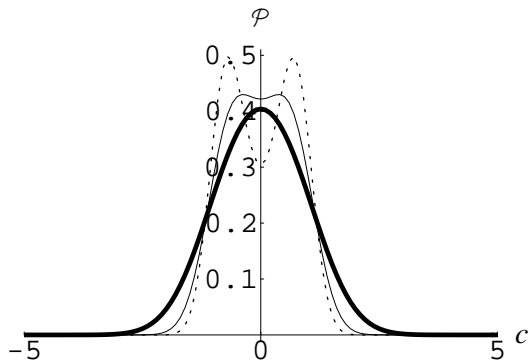}
\caption{\label{figprob-dens} The probability density distribution
of $c$ variable for $J=k_e=1$, $\nu=0.25$, $\chi=2$,
$\sigma=10^5$ and three different values of the temperature:
$T=0.05$ (dashed curve), $T=0.1$ (thin curve), $T=0.3$ (thick
curve)}
\end{figure}

There are two areas in the parameter space $\left(\beta,\xi\right)$
where the probability density (\ref{reduced}) as a function of the
curvature parameter $c$ behaves qualitatively different as is shown in
Fig. \ref{figphase}. These two areas are separated by the curve
\begin{equation}\label{sep-curve}
\xi=\frac{\beta}{\pi}\,\,\frac{2+\,\pi\, \left(I_1(\beta)+\,{\bf
L}_{1}(\beta)\right)}{\beta\,\left(I_0(\beta)+\,{\bf
L}_0(\beta)\right)-2\,\left(I_1(\beta)+ \,{\bf L}_1(\beta)\right)}
\end{equation}
where $I_n(\beta)$ is the Bessel function of imaginary argument and
${\bf L}_n(\beta)$ is the Struve function \cite{abr}. Below this
curve ({\it i.e.} in the unshaded area of the phase diagram
presented in Fig. \ref{figphase}) the distribution  is single-modal.
The probability density (\ref{reduced}) in this case has a maximum
at $c=0$ (see Fig. \ref{figprob-dens} when $T=0.3$). This means that
the most probable conformation state of the filament is a circle.
Above the curve (\ref{sep-curve}) ({\it i.e.} in the shaded area of
Fig. \ref{figphase}) the probability distribution is bimodal. The
function (\ref{reduced}) has two equivalent maxima $\pm c_m$ (see
Fig.\ref{figprob-dens} when $T=0.05$ and $T=0.1$). In this case the
most probable state is an elliptically deformed filament.
Two maxima correspond to two mutually orthogonal directions
of elongation.

It is straightforward to obtain that in the vicinity of the curve
given by Eq. (\ref{sep-curve}) the most probable value of the chain
curvature is determined by the expression
$c_m=\gamma(\xi)\,\sqrt{T_c-T}$ where $T_c(\xi)$ is a critical
temperature ({\it i.e.} the solution of Eq. (\ref{sep-curve})) and
$\gamma(\xi)$ is some coefficient. Thus, in the framework of our
Ansatz (\ref{fourier-reduce1}) the transition of the filament in the
fluctuating environment to a non-circular conformation  may be
considered as a noise-induced phase transition of the second kind.

By using the probability distribution (\ref{reduced}) and the
expression (\ref{aspherity-anal}) for the filament aspherity, one
can calculate the equilibrium value of the normalized aspherity
\begin{equation}\label{asp-equi}
\frac{\langle A\rangle}{\langle A_0\rangle} =\frac{M_1(T)}{M_1(0)},
\end{equation}
and the equilibrium value of the relative standard deviation of the
aspherity
\begin{equation}\label{apsh-dev-anal}
\frac{\sqrt{\langle\left(\Delta A\right)^2\rangle}}{\langle
A\rangle} = \sqrt{\frac{M_2(T)}{M_1^2(T)}-1},
\end{equation}
where $M_n(T)$ are the moments defined as
\begin{equation}\label{kappanu}
M_{n}(T)=\int\limits_{-1}^{1} {\cal
P}\left(c\right)\,|c|^n\,d\,c,\qquad n=1,2,...
\end{equation}


\section{Numerical studies}

The dynamics of the filament is described by the Schr{\"o}dinger
equations
\begin{equation}\label{dyneqpsi}
i\,\frac{d}{dt}\psi_l=
-\frac{\partial\, H}{\partial\,\psi^*_l}
\end{equation}
and the Langevin equations
\begin{equation}\label{dyneqr}
\eta\,\frac{d}{dt}\vec{r}_l=-\frac{\partial
H}{\partial\,\vec{r}_l}+\vec{R}_l(t).
\end{equation}
with the Hamiltonian $H$ being defined by Eq. (\ref{hamilt}).
Thus the conformational dynamics is considered in an overdamped
regime with $\eta$ being the friction coefficient. In accordance
with the fluctuation-dissipation theorem the standard deviation
$D$ is proportional to the temperature $T$, thus ~$D=\eta~T$.

The set of stochastic differential Eqs. (\ref{dyneqpsi}),
(\ref{dyneqr}) is solved numerically by the use of an implicit
Euler method with $\alpha=0.5$ as implicitness parameter, which is
the trapezoidal rule. The integration of the stochastic term is
done by using the strong Taylor scheme of first order described in
Ref. \cite{kloeden}. The time step chosen for running the
simulations was $\Delta t=10^{-2}$. To verify the precision of the
results we compared with data obtained for different time steps.
The value of the position and charge of $l$'th particle at time
$t_n=n\Delta t$ is denoted $Z^l_n=(x_l,y_l,\varphi_l)_n$,
$F(\vec{Z}_{n})$ denotes the deterministic part of Eqs.
(\ref{dyneqpsi}), (\ref{dyneqr}) and $W_n = (X_l(t_n),Y_l(t_n),0)$
is the corresponding component of the white noise. Thus the
numerical scheme becomes
\begin{eqnarray}
Z^l_{n+1} =Z^l_n + [\alpha F(\vec{Z}_{n+1}) +
(1-\alpha)F(\vec{Z}_{n})] \Delta t \nonumber\\
+ W_n \sqrt{\Delta t}.
\end{eqnarray}
The above system of nonlinear equations is implicit and it is
solved by a hybrid method provided by the minpack FORTRAN library
and the random numbers present in the white noise are generated by
the ranlib library, both accesible from the ``netlib'' repository
in Ref \cite{netlib}.

In our paper we were mostly concerned with the role of
thermal fluctuations in the process of the shape transformations.
Therefore in our simulations we have chosen a set of parameters
for which in the zero-temperature limit the most energetically
favorable state is an elliptically deformed filament. Without loss
of generality, in this section we show the results of the numerical
simulations produced for $L=36$ units and charge density $\nu=0.5$.
A system of this size provides
a clear visualization of the properties of the model and does not
demand too heavy computational time. The systems of the same size
but with smaller values of the charge density (however, still
inside the area where the elliptic state is stable) in the
presence of thermal fluctuations require more time to reach an
equilibrium state. As initial condition for electric
charge density $\psi_l$ we use the same magnitude at all points,
corresponding to an equally distributed charge density. Initially,
all the lattice points were symmetrically distributed along the
circle of an appropriate radius.  In what follows we have chosen
the damping coefficient $\eta$ and the bending rigidity $k$ equal
to unity. Moderate changes in these parameters do not
modify significantly the dynamics of the system. To avoid  big
stretching of the nearest bonds of the chain we fix the parameter
$\sigma=10^5$. We considered both the cases of the hardening and
of the softening electron-curvature interaction.


\subsection{The case of hardening charge-curvature interaction}

\begin{figure*}[ht]
\centering
\includegraphics[width=15.cm]{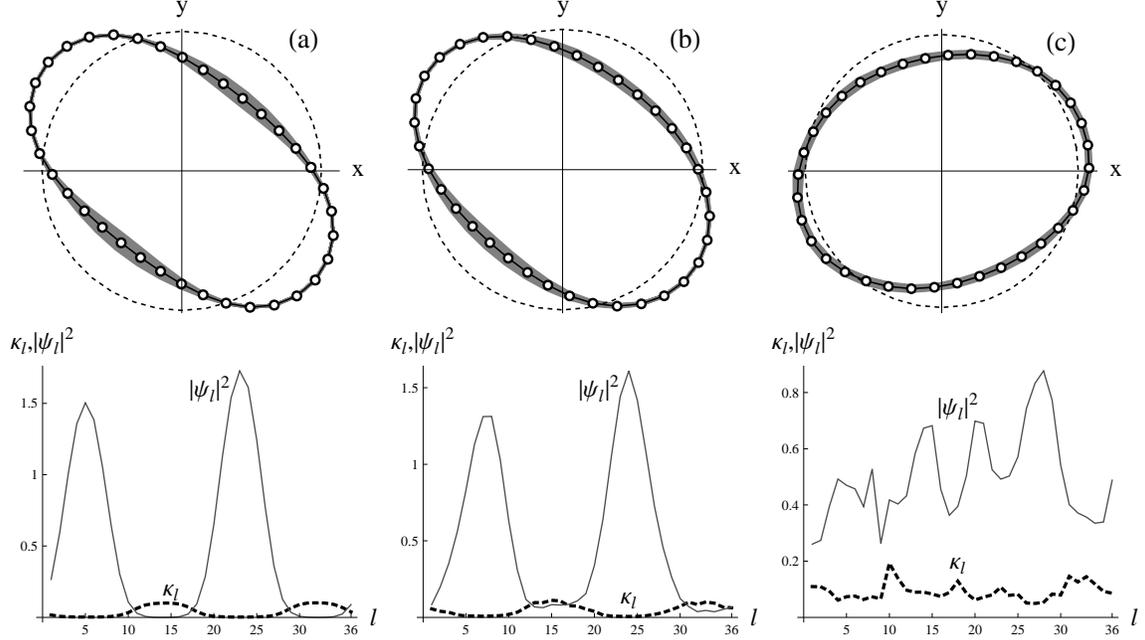}
\caption{The top panel shows the equilibrium shape of the chain
(solid line) against the initial circular shape (dashed line). The
bottom panel shows the charge distribution $|\psi|^2$ (solid line) and
curvature $\kappa$ (dashed line) for different noise intensity:
$D=0.01$ (a), $D=0.2$ (b), and $D=0.4$ (c) in the case of
hardening, $\nu=0.5$, $\chi=-4$, $\sigma=10^5$ and $J=0.25$ at time $t=4000$. The
grey shadow represents the local charge density in the chain.
\label{hard_nu050_chi4_g32shape_circles}}
\end{figure*}

\begin{figure}[h]
\centering
\includegraphics[width=8.cm]{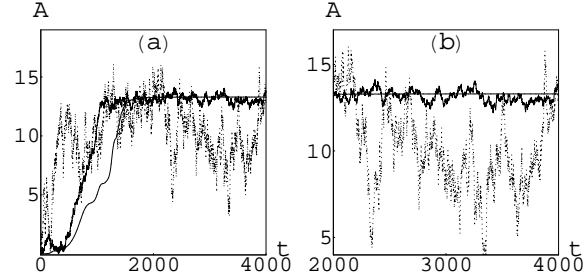}
\caption{Aspherity versus time for different noise intensity,
$D=0$ (thin line), $D=0.05$ (thick line) and $D=0.25$ (dotted
line). The parameters used are $\nu=0.5$, $\chi=-4$, $\sigma=10^5$ and $J=0.25$.
(a) Full time simulation $0\le t \le 4000$, (b) Detailed behavior
after transient time $2000\le t \le 4000$
\label{hard_nu050_chi4_g32asp_time}}
\end{figure}

Typical final shapes of the filament for three different values of
the noise intensity $D=0.01$, $D=0.1$ and $D=0.2$ are shown in
Fig. \ref{hard_nu050_chi4_g32shape_circles} where the averaged
aspherity decreases as the noise increases. Fig.
\ref{hard_nu050_chi4_g32asp_time} shows the time evolution of the
aspherity $A$ for three different values of the noise
$D$. The left panel shows the overall behavior of this quantity
(including transient processes) while the right panel presents its
steady state evolution. As it is seen from Fig.
\ref{hard_nu050_chi4_g32asp_time} the thermal fluctuations excite
the dynamics of the system and facilitate the transition to an
anisotropic state. Clearly the transient period shortens when the
temperature increases.

\begin{figure}[h]
\centering
\includegraphics[width=8.cm]{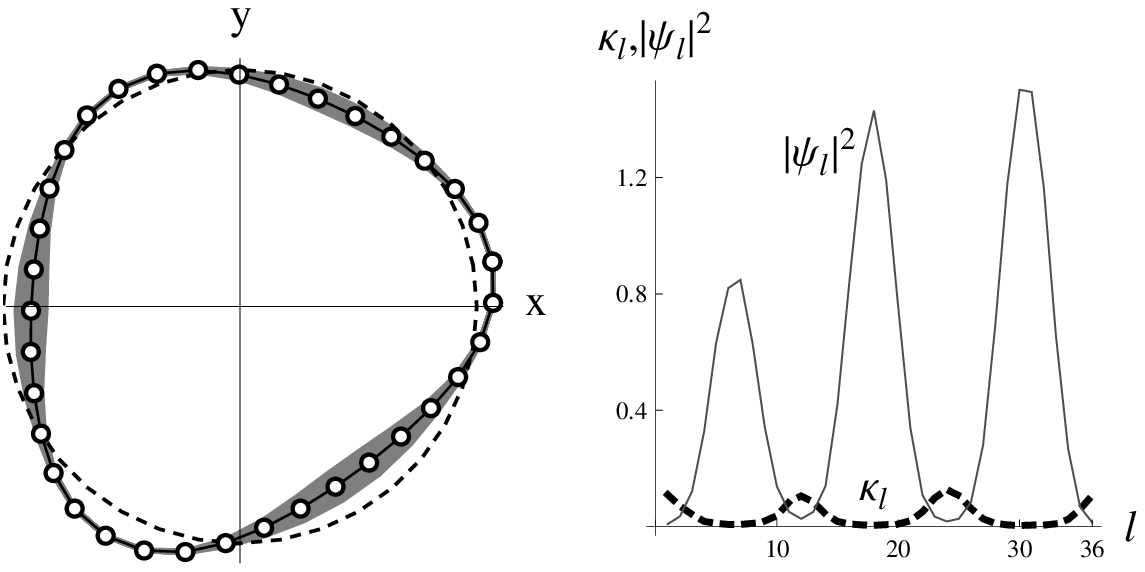}
\caption{The intermediate shape of the filament at time $t=500$
(left panel) and the corresponding charge distribution $|\psi|^2$ (solid
line) and curvature $\kappa$ (dashed line) along the chain (right
panel) for $D=0.01$. The parameters used are the same as in Fig.
\ref{hard_nu050_chi4_g32shape_circles}
\label{hard_nu050_chi4_g32shape_triangle}}
\end{figure}

It is remarkable that in the case of weak noise, $D=0.01$, in the
aspherity time evolution there exists a small plateau for
$600<t<1000$ (see Fig. \ref{hard_nu050_chi4_g32asp_time}). This
plateau corresponds to an intermediate conformation of the
filament when it takes a triangular shape (see Fig.
\ref{hard_nu050_chi4_g32shape_triangle}).

\begin{figure}[h]
\centering
\includegraphics[width=8.cm]{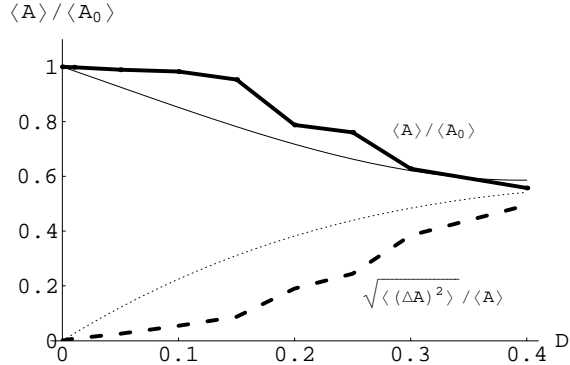}
\caption{The normalized mean value of the aspherity $\langle
A\rangle/\langle A_0\rangle$ (thick solid line) and its
relative standard deviation $\sqrt{\langle\left(\Delta
A\right)^2\rangle}/\langle A\rangle$ (thick dashed line)
versus noise from the simulations and as stationary solution of the
Fokker-Planck equation (thin solid line and dotted line
respectively) in the case of hardening, $\nu=0.5$, $\chi=-4$,
$\sigma=10^5$ and $J=0.25$.
\label{hard_nu050_chi4_g32asp_D_analytic}}
\end{figure}

The mean value of the saturation aspherity $\langle A\rangle$ after
transient time $t_1$ which we define as
\begin{equation}\label{meanasp}
 \langle A \rangle=\frac{1}{t_2-t_1}\,\int\limits_{t_1}^{t_2}\,A(t)\,dt
\end{equation}
decreases as the temperature increases (see Fig.
\ref{hard_nu050_chi4_g32asp_D_analytic}) and the filament takes on
a less anisotropic shape while the relative standard deviation
\begin{equation}\label{relvar}\frac{\sqrt{\langle
\left(\Delta A\right)^2\rangle}}{\langle A\rangle}
\end{equation}
is an increasing function of temperature. Here the aspherity variance
is given by the expression
\begin{equation}\label{meanaspdev}
 \langle \left(\Delta A\right)^2\rangle = \frac{1}{t_2-t_1}\,
 \int\limits_{t_1}^{t_2}\,\left(A(t)-\langle A \rangle\right)^2 \,dt.
\end{equation}
The shape of the aggregate is well-defined when
$\sqrt{\langle\left(\Delta A\right)^2\rangle}/\langle A\rangle\ll 1$.
There is a critical value of the noise intensity
$D_{cr}$ when the relative standard deviation reaches the value $1/2$.
Then for $D\,>\,D_{cr}$ the shape fluctuations are so strong that
they become chaotic. For $\nu=0.5$ this
value is approximately $D_{cr}\approx 0.4$.

Thus one can conclude that in the case of hardening charge-bending
interaction the mean-field approach introduced in \cite{gaid-conf}
works well in the weak-noise limit:

(i) The ellipse-like conformation is the equilibrium state of the
shape evolution.

(ii) In the equilibrium state the charge distribution is
nonuniform along the chain and the charge is concentrated in the
places where the filament is more flat.

(iii) The triangular conformation of the filament exists but it is
a metastable state.

Strong noise ({\it i.e.} high temperature) changes qualitatively
the conformational dynamics keeping the shape of the filament
almost circular, while in the zero-temperature limit it takes on
an elliptic shape.

The corresponding quantities are shown in Fig.
\ref{hard_nu050_chi4_g32asp_D_analytic}. Comparing these relations
with the numerical data, also presented in this figure, one can
conclude that there is a qualitative agreement between the
analytical results obtained in the framework of the simple Ansatz
(\ref{fourier-reduce1}) and the results of numerical simulations.

\subsection{The case of softening charge-curvature interaction}

By carrying out the simulations described for the hardening
charge-bending interaction ($\chi<0$) we took into account that
the presence of charge increases the bending rigidity locally and
the large bending deformations do not occur. Therefore there was
no necessity to introduce any bending energy penalty and in Eq.
(\ref{potbend}) we put $\kappa_{max}\rightarrow\infty$. However,
the filaments with softening charge-bending interaction are much
more flexible and large deformations with the bending angle
$\alpha_l>\pi/2$ appear rather easily. To avoid excessive bending
we carried out our numerical simulations under the assumption that
the maximum possible bending angle $\alpha_{max}$ is equal to
$\pi/2$. This means that we let $\kappa_{max}=\sqrt{2}$ in Eq.
(\ref{potbend}).

\begin{figure}[h]
\centering
\includegraphics[width=8.cm]{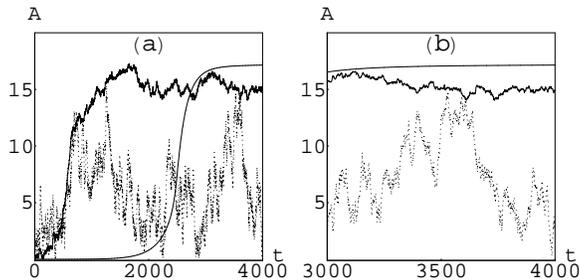}
\caption{Same as  Fig. \ref{hard_nu050_chi4_g32asp_time} for the
case of softening, $\nu=0.29$, $\chi=2$, $\sigma=10^5$ and $J=0.22$. (a) Full
time simulation $0\le t \le 4000$, (b) Detailed behavior after
transient time $3000\le t \le 4000$
\label{soft_nu029_chi2_g18asp_time}}
\end{figure}

\begin{figure}[h]
\centering
\includegraphics[width=8.cm]{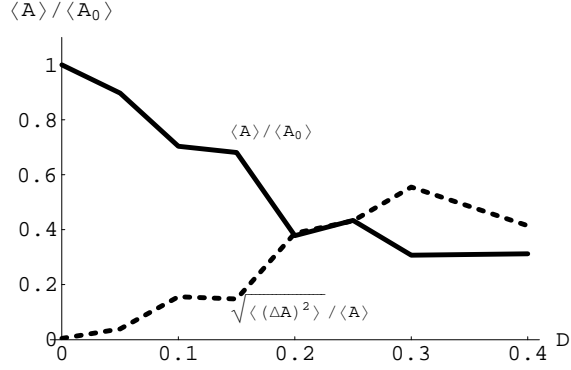}
\caption{The normalized mean value of the aspherity $\langle
A\rangle/\langle A_0\rangle$ (solid line) and the relative
standard deviation
$\sqrt{\langle\left(\Delta A\right)^2\rangle}/\langle A\rangle$
(dashed line) versus noise for the case of softening with
$\nu=0.29$, $\chi=2$, $\sigma=10^5$ and $J=0.22$.
\label{soft_nu029_chi2_g18_asp_D}}
\end{figure}

\begin{figure}[h]
\centering
\includegraphics[width=8.cm]{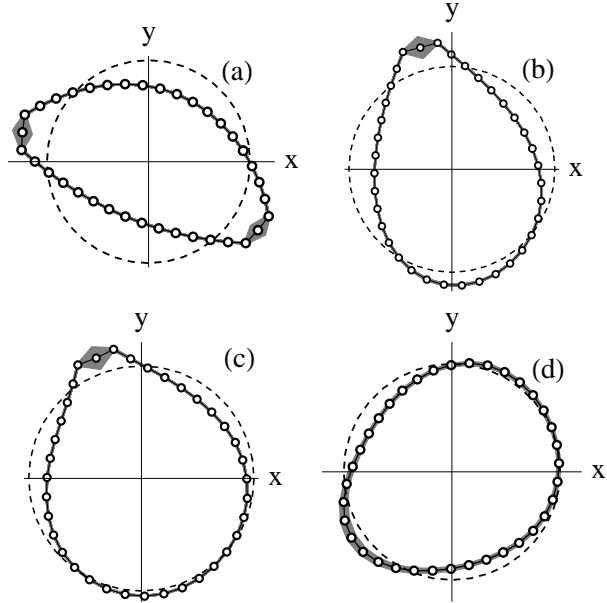}
\caption{Equilibrium configurations of the chain in the case of
softening, $\nu=0.29$, $\chi=2$, $\sigma=10^5$ and $J=0.22$, for noise
intensities: (a) $D=0.05$, (b) $D=0.10$, (c) $D=0.20$ and (d)
$D=0.30$ (solid line) against the initial circular conformation
(dashed line). The grey shadow represents the charge density.
 \label{soft_nu029_chi2_g18shape_circles}}
\end{figure}

\begin{figure}[h]
\centering
\includegraphics[width=8.cm]{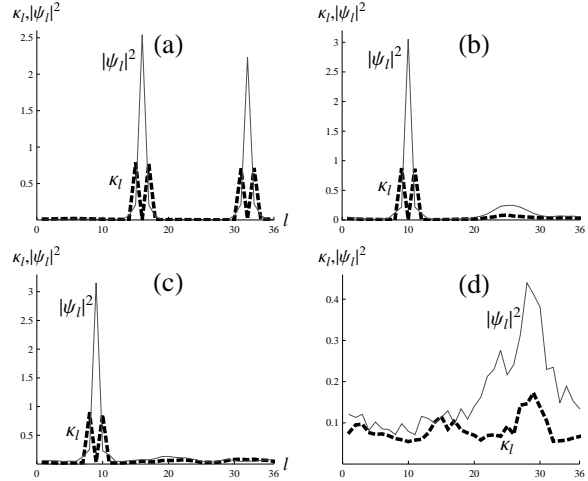}
\caption{Equilibrium charge distribution $|\psi|^2$ (solid line) and
curvature $\kappa$ (dashed line) along the chain for the same
parameters as in Fig. \ref{soft_nu029_chi2_g18shape_circles}.
\label{soft_nu029_chi2_g18curv_wnorm}}
\end{figure}

Comparing the time evolution and the mean value of the aspherity
in the case of softening charge-bending interaction presented in
Fig. \ref{soft_nu029_chi2_g18asp_time} and Fig.
\ref{soft_nu029_chi2_g18_asp_D} with the case of hardening
charge-bending interaction (Fig. \ref{hard_nu050_chi4_g32asp_time}
and  Fig. \ref{hard_nu050_chi4_g32asp_D_analytic}) we see that
they are qualitatively similar. However, the equilibrium shapes of
the filaments (Fig. \ref{soft_nu029_chi2_g18shape_circles}) and
the charge distribution (Fig. \ref{soft_nu029_chi2_g18curv_wnorm})
after transient time $t=4000$ differ drastically. The softening
charge-bending interaction initiates {\it kink formation} while
smooth shapes of filaments are characteristic for the case of
hardening charge-bending interaction. Almost all the charge is
concentrated now in the areas of the kinks. It is interesting to
notice that small and moderate thermal fluctuations facilitate the
kink formation while for sufficiently high magnitudes of the noise
intensity filaments takes on almost circular shape and the charge
distribution is more uniform along the filament (see Fig.
\ref{soft_nu029_chi2_g18shape_circles}(d) and Fig.
\ref{soft_nu029_chi2_g18curv_wnorm}(d)).

We carried out a few runs of numerical simulations for  a
set of parameters which satisfy the inequality (\ref{A5}). We found
out that for small enough $V$ the filament shape behavior and
aspherity behavior are essentially the same as in the case of
no-interaction between charges. More detailed studies of effects of
charge-charge interactions will be published elsewhere.


\section{Conclusions and discussion}
In this paper, we have investigated the role of thermal
fluctuations on the charge-induced conformational
transformations of closed semiflexible molecular chains. We have
found that the results obtained in the mean-field approach
\cite{gaid-conf} are rather robust in the systems  where the
presence of charge {\it hardens} the local chain stiffness, the
charge-curvature interaction counteracts the collapse of the chain
and  the mean-field picture survives. In the presence of white
noise when the charge density and/or the strength of the
charge-curvature coupling exceed a threshold value, the spatially
uniform distribution of the charge along the chain and the
circular, cylindrically symmetric shape of the chain  become
unstable. In this case the equilibrium state of the system is
characterized by a spatially nonuniform charge distribution along
the chain which takes on an ellipse-like form. The transition to
an anisotropic spatially nonuniform conformation is analogous to
the phase transition of the second kind in the condensed matter
physics.

In the case of hardening charge-bending interaction the charge and
curvature distribution along  the filament are smooth while for
softening charge-bending interaction there spontaneously create {\it
kinks} where the smooth filament structure is disrupted. Almost
total amount of excess charge is concentrated in the vicinity of the
kinks.


\begin{acknowledgments}
Yu.\,B. G. and C. G. thank Department of Mathematics, Technical
University of Denmark for hospitality. C. G. acknowledges the
project MTM2007-62186 granted by the Spanish Ministerio de
Educaci\'on y Ciencia. Yu.\,B. G. acknowledges also support from
the Special Program of Department of Physics and Astronomy of the
National Academy of Sciences of Ukraine and
Civilingeni{\o}r Frederik Christiansens ALMENNYTTIGE FOND.  M.\,P. S.
acknowledges support from the European Union through the Network
of Excellence BioSim, LSHB-CT-2004-005137. We acknowledge
financial support from the Danish Center for Applied Mathematics
and Mechanics (DCAMM), International Graduate Research School,
contract number 646-06-004, the Danish Agency for Science,
Technology and Innovation. Finally this work received funding from
the Mathematical Network in Modelling, Estimation and Control of
Biotechnological Systems (MECOBS), contract number 274-05-0589,
the Danish Research Agency for Technology and Production.
\end{acknowledgments}


\bibliographystyle{apsrev}

\bibliography{biblio}

\end{document}